\documentclass[fleqn,12pt,twoside]{article}
\usepackage{espcrc1}
\usepackage{epsfig}

\usepackage{graphicx}
\usepackage[figuresright]{rotating}

\def\epol{\vec{e}p}

\def\Dpol{\vec{e}\vec{D}}
\def\Npol{\vec{e}\vec{N}}
\def\as{\alpha_s}

\newcommand{\AmS}{{\protect\the\textfont2
  A\kern-.1667em\lower.5ex\hbox{M}\kern-.125emS}}

\title{HERA-3\footnote{Contribution to the Proceedings
              of QCD-N'02 Workshop, Ferrara, Italy, 
              3-6 April 2002}} 

\author{S. Levonian\address{DESY\\ 
        Notkestrasse 85, 22607 Hamburg, Germany \\
        E-mail: levonian@mail.desy.de}}%

\begin{document}

\maketitle

\begin{abstract}
A brief review is given on the physics potential of HERA beyond
the presently approved programme. Questions of particular interest
include QCD in a regime of weak coupling and high partonic densities
achieved in $eA$ collisions and spin physics with polarized 
colliding beams. Those two distinct programmes can be naturally
merged and even substantiated by using polarized and unpolarized deuterons
at the starting phase of the HERA-3 run. Basic machine implications and 
detector requirements are also discussed.
 
\end{abstract}

\section{INTRODUCTION}

HERA, an electron-proton facility at DESY, serves two collider and two 
fixed target experiments.
H1 and ZEUS explore high centre of mass energy up to 318 GeV to investigate a
structure of the proton at smallest achivable dimensions ${\cal O}(10^{-18}$m).
HERMES focuses on spin physics using $27.6$ GeV polarized lepton beam with
$\langle P_b\rangle \approx 55\%$
and different polarized targets. HERA-B studies charm and beauty sector
of the Standard Model by utilizing 920 GeV proton beam interacting with 
internal wire target.

First running period, so called HERA-1, 
was completed in year 2000 and has delivered 120 pb$^{-1}$ per 
collider experiment. Presently, after the luminosity upgrade, data taking
has started at HERA-2 with the aim being to collect polarized $\epol$
scattering data of $1$ fb$^{-1}$. To fully exploit the potential of this
unique facility it is proposed to extend its running for another five years
available between HERA-2 and TESLA~\cite{TESLA} -- 
a major future international HEP project.
This third running period, termed here HERA-3, would provide yet deeper insight
into fundamental QCD questions, which is not possible in present HERA 
configuration.

\section{QCD PHYSICS AT HERA-1 AND PROSPECTS FOR HERA-2}

Before discussing the future of lepton-nucleon scattering
let see what did we learn at HERA and where will we be 
by the year 2006/7, the end of the HERA-2 data taking period.
 In this context we focus on QCD aspects
and the structure of the nucleons.

\begin{figure}[htb]
  \begin{picture}(16.7,8)(0.1,0) 
    \put(0.0,0.0){\epsfig{file=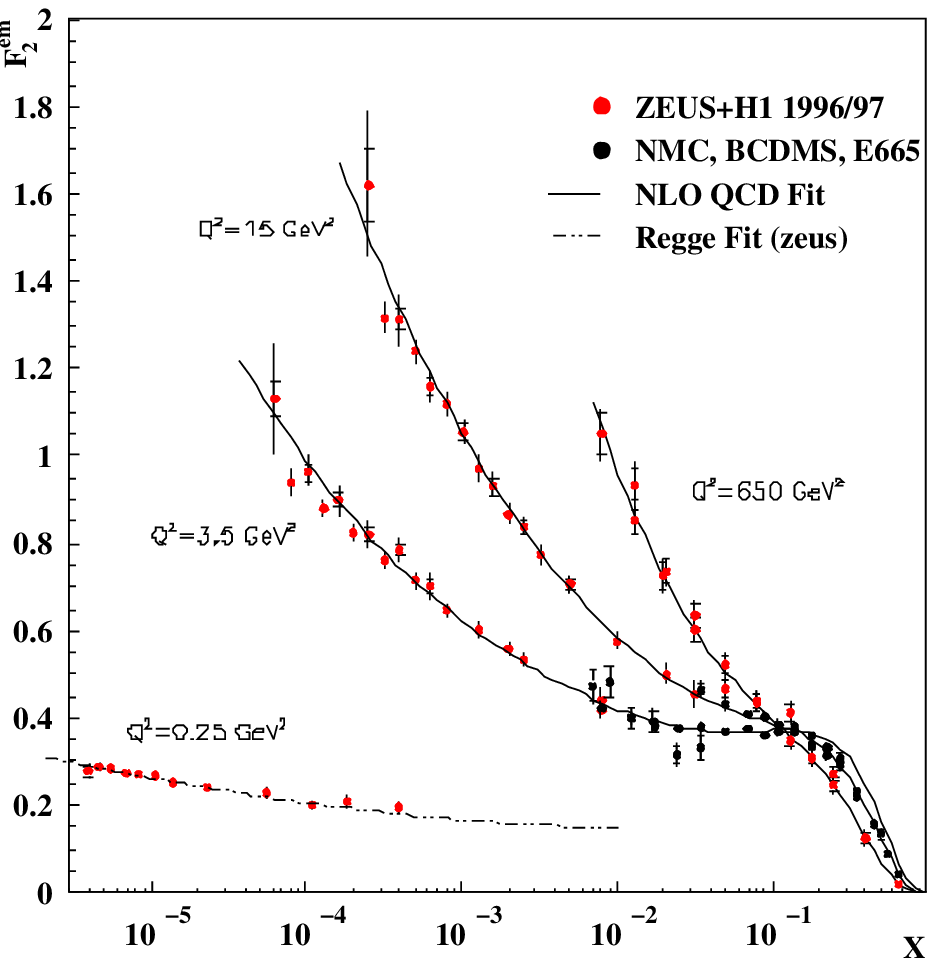,width=7.7cm}}
    \put(8.2,-0.4){\epsfig{file=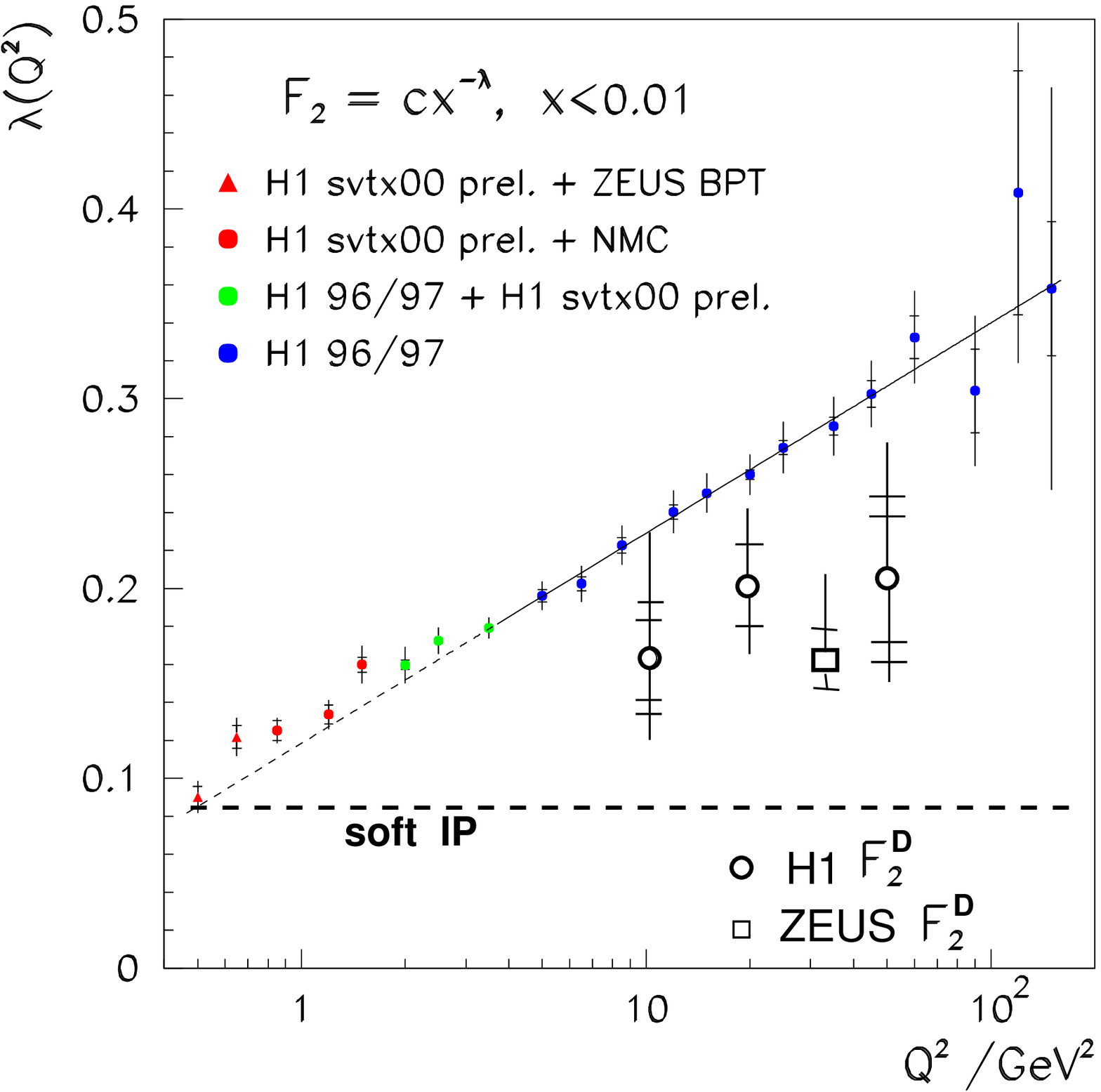,width=8.2cm}}
    \put(1.0,7.0){(a)} \put(15.0,7.0){(b)}
  \end{picture}
  \caption{a) Proton structure function at low $x$ and different $Q^2$;
           b) The low $x$ behavior of the proton structure function 
              in inclusive and diffractive DIS.}
  \label{fig:1} 
\end{figure}

Low $x$ physics  and related to that diffractive DIS constitute 
a major impact of HERA data on QCD in its high energy limit 
($x \propto Q^2/s$). The discovery of the strong rise of the 
proton structure function $F_2$ at low $x$ has triggered a 
discussion about when and how this steep rise should slow down
due to parton saturation.
Fig~\ref{fig:1}a shows that NLO QCD using linear DGLAP evolution
equations is able to describe inclusive DIS data in whole
$x$ range down to very low $Q^2,~{\cal O}(2$GeV$^2)$. This suggests
the saturation regime is not yet reached at HERA. 

Still, since the low-$x$ behavior is governed by gluons one may hope 
first to see a sign of saturation in diffraction whose cross section
is proportional to the gluon density squared: 
$\sigma_{diff} \propto |xg(x)|^2$. Indeed, by comparing the energy
rise of the inclusive and diffractive DIS cross sections, parametrised
as $\sigma_{tot}(\gamma^*p) \sim W^{2 \lambda(Q^2)}$  
and $\sigma_{diff}(\gamma^*p) \sim W^{4 \lambda(Q^2)}$ respectively,
one can see some systematic shift between the two (Fig.~\ref{fig:1}b).
The error bars for diffractive points are still large, and one of the
tasks for HERA-2 is to improve precision of $F_2^D$ measurement and
to quantify this so far only qualitative statement.  

To summarize, there is a controversy about parton saturation at HERA,
both between theorists and experimentalists. It is believed that the
saturation scale~\cite{GLR,CGC}, 
$Q^2_{sat}=\as N_c \frac{dN_g}{dy}/\pi R^2$,
is too small in HERA-ep. One needs conditions when linear DGLAP evolution
equations break down unambiguously and the parton shadowing effects 
(or the processes of gluon recombination)
become apparent yet in perturbative domain 
($Q^2_{sat} \gg \Lambda_{QCD}^2, ~~~ \as(Q^2_{sat}) \ll 1$). 
In order to achieve this, gluon density should somehow be magnified. 
It is possible either by increasing energy,
$xg(x,Q^2) \sim s^{\lambda(Q^2)}$ (THERA~\cite{THERA} option), or
by using nuclear target instead of protons, $xg_A(x) \sim A^{1/3} xg_p(x)$ 
(HERA-3, EIC~\cite{EIC}).   
As an example, $eCa^{40}$ scattering at HERA-3 would be equivalent
in terms of gluon density to $1.8 {\rm TeV} (e) \times 0.9 {\rm TeV} (p)$
collider.

\begin{figure}[htb]
  \begin{picture}(16.7,7.5)(0.1,0)
    \put(0,0){\epsfig{file=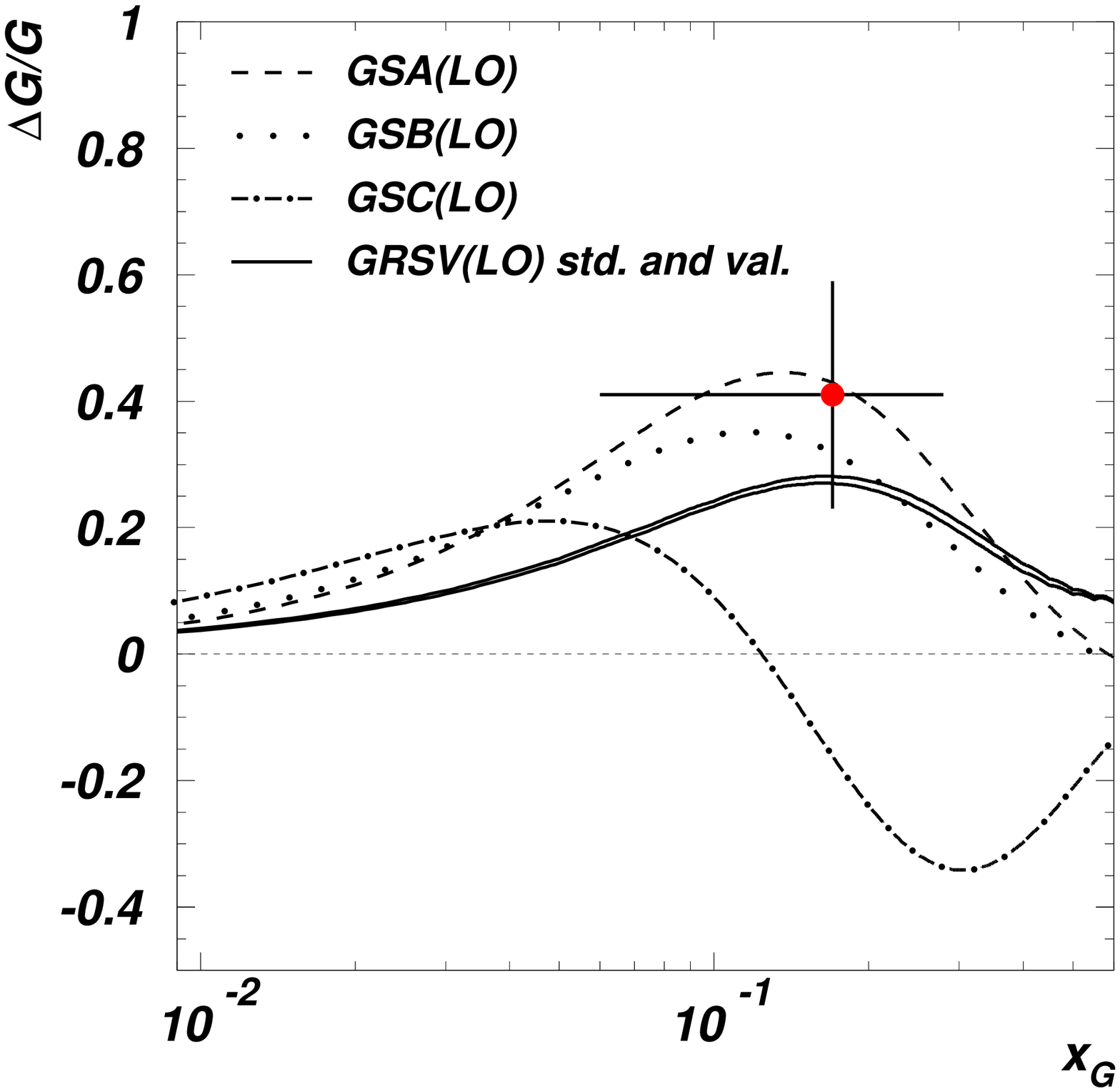,height=7.2cm}}
    \put(7.9,-0.1){\epsfig{file=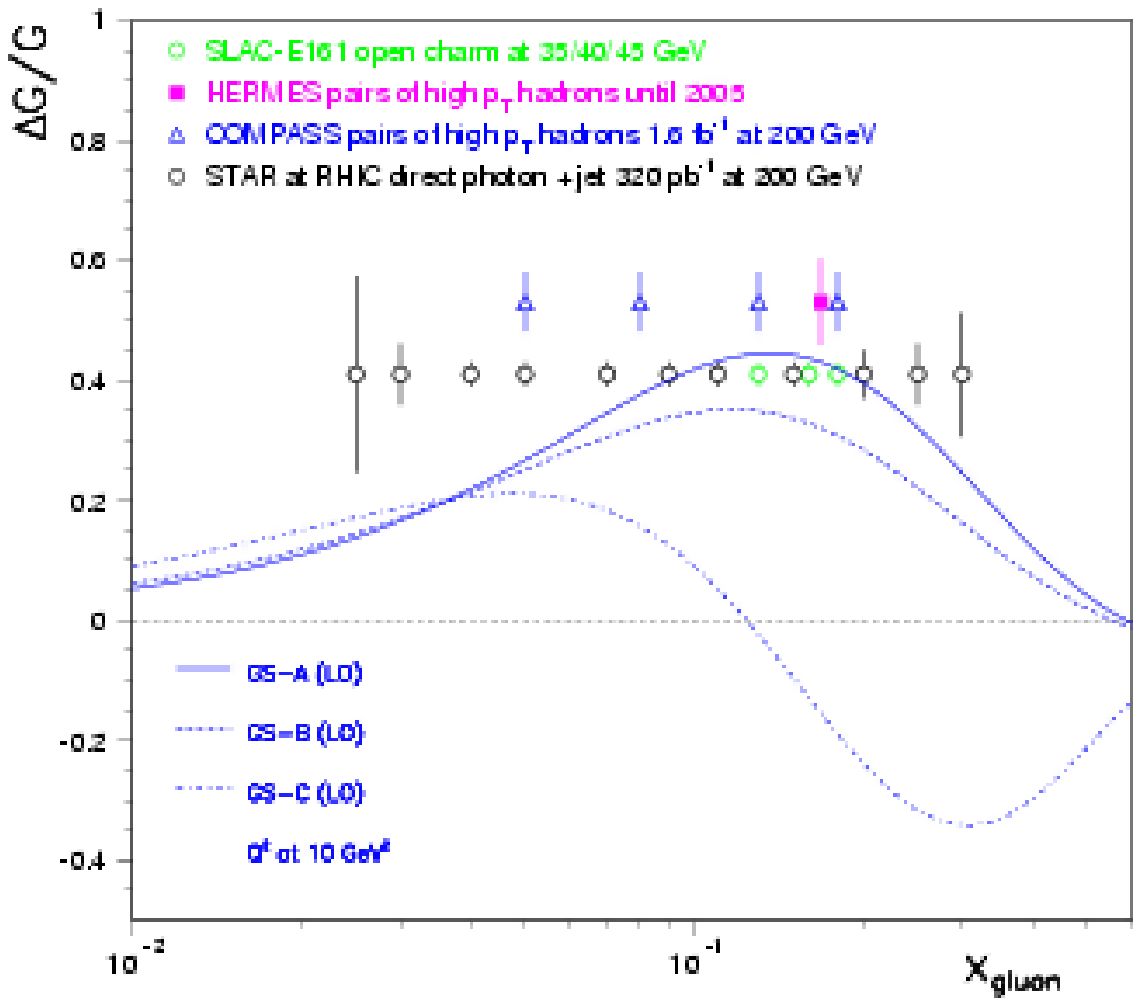,height=7.7cm}}
    \put(6.2,6.4){(a)} \put(15.5,6.4){(b)}
  \end{picture}
\caption{a) Gluon polarization as extracted by HERMES from 
            photoproduction of high $p_t$ hadrons~\cite{HERMES}.
         b) Projected precision of the gluon polarization 
            extraction in different experiments, by the end of 
            HERA-2~\cite{GvS}.}
\label{fig:2}
\end{figure}

Another important aspect of QCD is the spin structure of the nucleon.
How do the partons conspire to ensure the total spin of the nucleon is $1/2$?
As quarks cannot account for the total nucleon spin (the fact known
since 1988 as "spin crisis") it is important to measure other components: 
gluon polarization $\Delta G$ and orbital angular momenta $L_{q,g}$.
In that respect first determination of $\Delta G$ by HERMES shown in
Fig.~\ref{fig:2}a is one of the highlights of HERA-1, in spite of
its limited precision. Anticipated quality of $\Delta G/G$ extraction
by the end of HERA-2 in different fixed target experiments
is illustrated in Fig.~\ref{fig:2}b.

There are however two intrinsic problems with $\Delta G$ measurement
at fixed target: a) majority of gluons are concentrated below accessible 
$x$ range and b) strong model dependence of the result, as higher scales are
necessary for the reliable extraction of gluon polarization. 
Additional advantage of higher energies is that several methods can be
used to measure $\Delta G$: from scaling violation of $g_1(x,Q^2)$, 
from dijet events and - statistics permitting - from heavy
quarks ($\gamma g \to c\bar{c}$). 
This motivates the use of the collider mode for such kind of measurements.

\section{HERA-3 OPTIONS}

Two main possibilities for HERA-3 have been investigated at previous workshops:
lepton-nuclei collisions~\cite{WS1} 
and physics with polarized colliding beams~\cite{WS2}.
They were recently summarized in ~\cite{WS3}.
It has been shown that electron-deuteron scattering appears to be a natural
next step following the $ep$ running. It is not only a first part of $eA$
programme, but also enables $F_2^n$ measurements, which is important on its own. 
Moreover, due to the small anomalous magnetic moment of the deuteron
the polarization is much easier to achieve with deuteron beam than with
protons~\cite{pol_D}. Hence this option can unify 
"spin" and "nuclear" physics communities at HERA-3.
 
\subsection{Physics subjects}

To summarize, the physics objective for HERA-3 is to shed more light
on the partonic structure of the neutron, on the nucleon spin decomposition
(especially at low $x$ and at high $Q^2$ domains) and on the problem of
gluon saturation and confinement. In particular, is so called Colour Glass
Condensate~\cite{CGC} indeed a new form of matter in high energy limit of QCD?

Key measurements in high energy $eA$ collisions include the ratio 
$\sigma_{diff}/\sigma_{tot}$, which is predicted to approach the black body
limit of $1/2$ for large $A$. 
$A$-dependence of the vector meson production, $\sigma_{VM}(x,A)$,
is expected to change the behavior from colour transparency ($\sim \!A^{4/3}$) 
to colour opacity ($\sim \!A^{0.4}$) with decreasing $x$. 
Inclusive structure function should violate DGLAP: $F_2 \sim Q^2\ln (x_0/x)$.  
These measurements at HERA-3 will also be valuable to understand more
complicated heavy ion collisions and could be important for LHC.

For polarized beams one would measure $g_1$ at low $x$ where it is essentially
unknown. Structure function $g_5$ can be for the first time determined
from charged
current asymmetries. 
DVCS will give access to the off-diagonal parton distributions at low $x$. 
It is important to stress that the study of spin phenomena in this
new kinematic domain will be enhanced by the comprehensive reconstruction
of the final state possible in colliding beam experiments.

\subsection{Machine Aspects}

Typical measurements listed in previous subsection put definite requirements
on the machine parameters. In case of polarized beams scattering most essential
are high luminosity, ${\cal L} \approx 500$ pb$^{-1}$, and beam polarization.
This is because one typically measures very small asymmetries at low $x$, or
low cross sections  at high $Q^2$. In case of proton beam one would need
spin rotators, flattening snakes and at least four Siberian snakes to maintain
polarization. Hence deuteron option looks very attractive, as there is good hope
to use transverse RF dipoles to rotate and stabilize longitudinal
polarization~\cite{pol_D} and thus to avoid complicated Siberian snake scheme.
An open question is a high quality source of polarized deuterons.

In contrast, expected effects and measured cross sections are large in $eA$
collisions. Hence moderate luminosity is sufficient: ${\cal L}\!\cdot\! A\approx 10$
pb$^{-1}$/ion. Low beam divergence is appreciated to minimize the $p_t$
smearing for elastically scattered and spectator nucleons.
Most complicated ingredient for this programme is electron cooling, which is
required for $A>4$ nuclei in order to keep an acceptable life time of the
ion beam. Another interesting question which deserves additional studies
is a possibility to fill and circulate different $A$ bunches simultaneously. 
This would reduce systematics in all measured $A$-dependencies drastically. 
\begin{figure}[htb]
  \begin{picture}(16.7,7.0)(0.1,0)
    \put(0.0,0.0){\epsfig{file=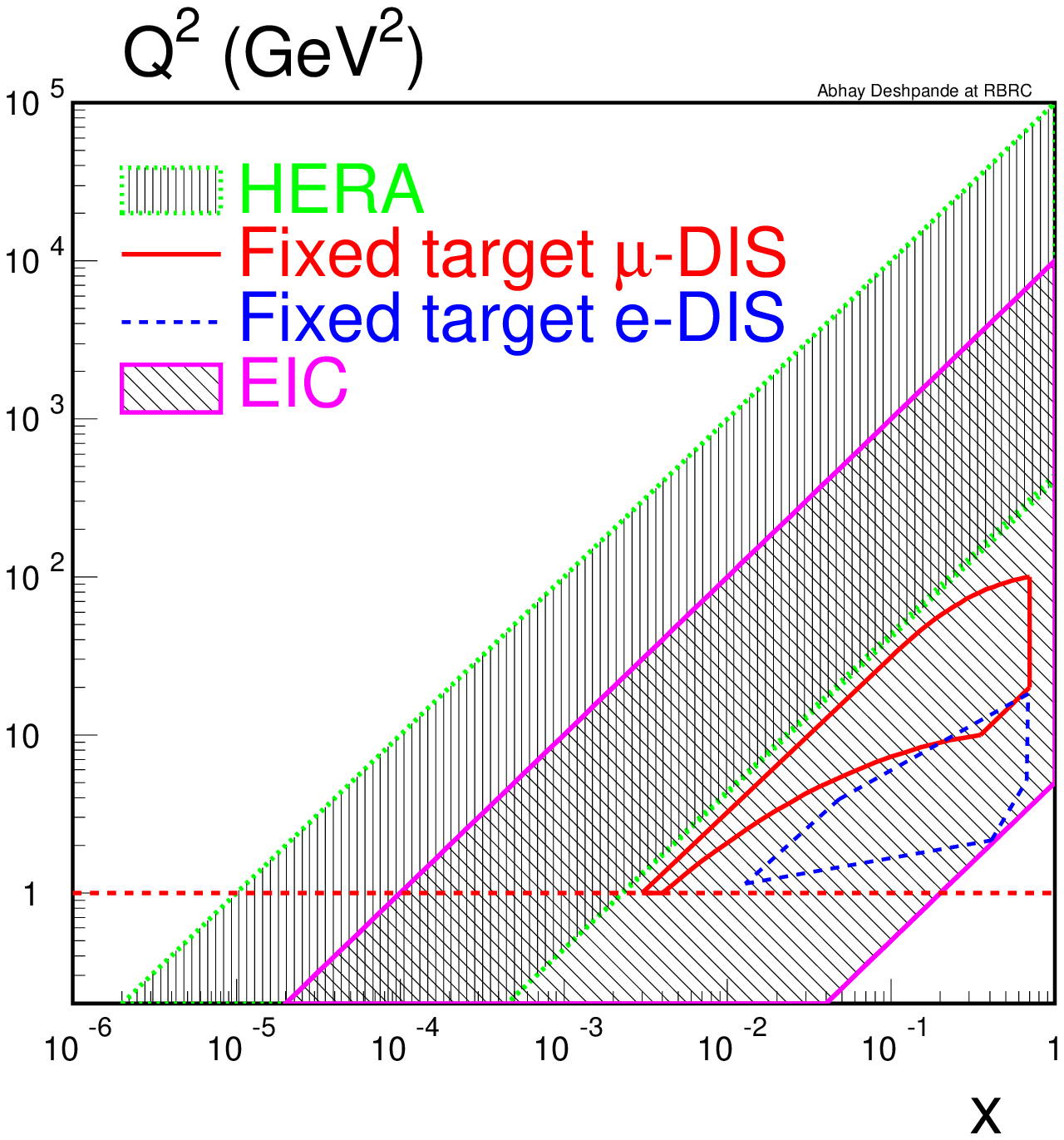,height=6.9cm,
    bbllx=100pt,bblly=240pt,bburx=470pt,bbury=620pt,clip= }}
    \put(7.7,-0.5){\epsfig{file=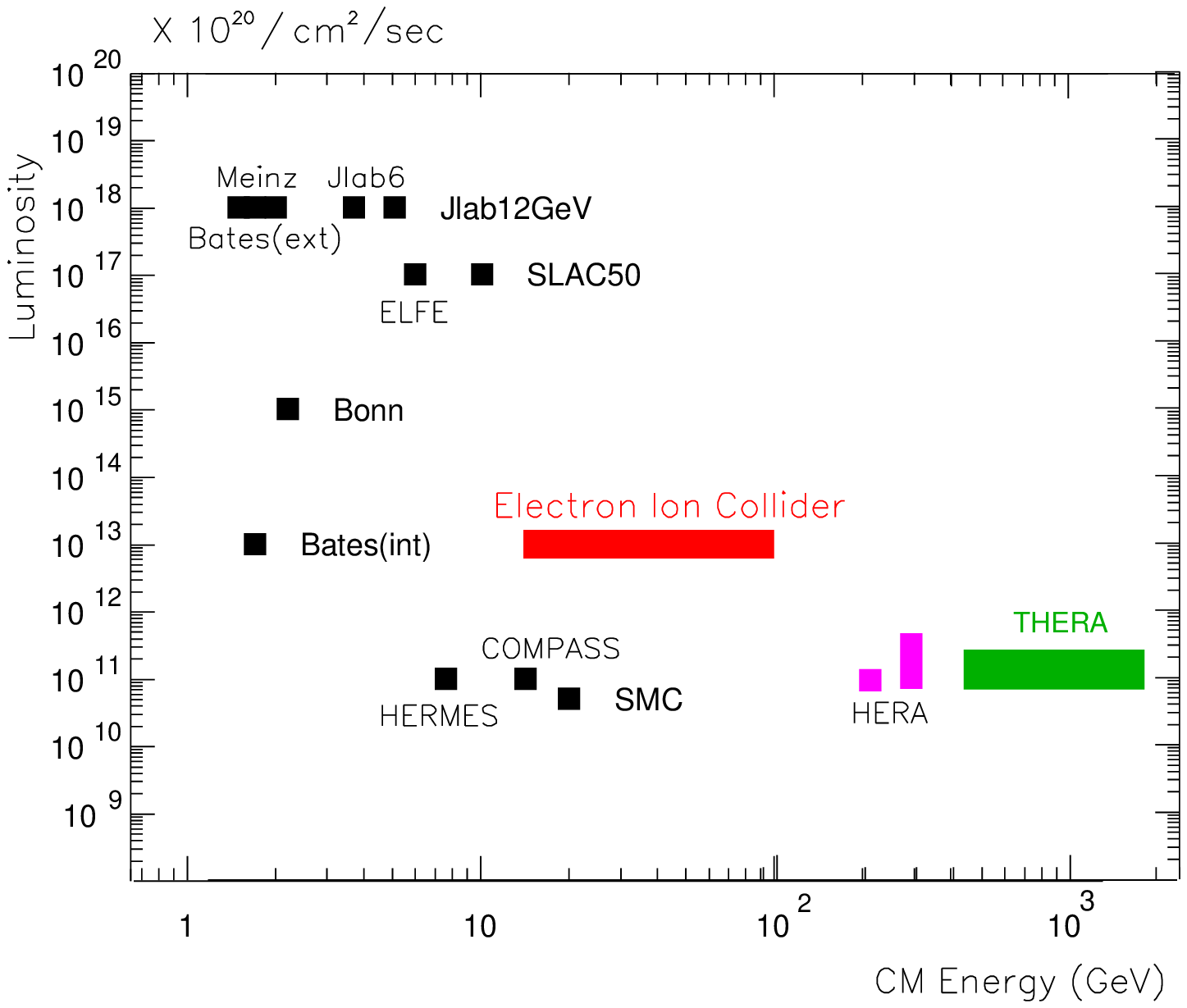,height=7.1cm}}
  \end{picture}
\caption{Comparison of the existing and proposed $eA$ and  
         polarized $\Npol$ experiments. Two entries for HERA
         correspond to $eA$ and $ep$ configurations respectively.}
\label{fig:3}
\end{figure}

In Fig.~\ref{fig:3} HERA is compared to other facilities and in particular
to the EIC~\cite{EIC}. The two projects are complementary
both in the covered phase space and in the planned commissioning time. Although
the EIC is superior in terms of anticipated luminosity it can use earlier
HERA-3 results to tune its physics programme.

\subsection{Detector Requirements}

Several options are possible of the experimental setup at HERA-3:
\begin{itemize}
  \item Retain existing setup: two colliding and one fixed target 
        experiment (H1, ZEUS, HERMES). 
        This in addition would allow to continue fixed target spin
        programme at HERA.  
  \item Only one general purpose collider detector. 
        This helps saving resources and concentrating efforts,
        and is also simpler for the machine tuning and stability.
  \item Two collider detectors optimized for different physics:
        low $x$ oriented -- high acceptance, $p/n$-tagging capabilities, 
        small beam divergence, and spin physics oriented --
        high ${\cal L}$, polarization, large statistics (stronger focusing,
        no beam-line detectors).   
\end{itemize}
Current evaluations show that main components of the present
H1 and ZEUS detectors could be reused for HERA-3. Some modifications
in central tracking and electronics may be necessary.
For $eA$ programme an essential requirement is an efficient nucleon tagging
capability. To study coherent diffraction $eA \to eAX$ 
large acceptance H1 VFPS stations~\cite{VFPS} at $z=220$m from 
interaction region can be used. The existing Forward Neutron Calorimeter~\cite{FNC}
at $z=107$m is important to distinguish central and peripheral $eA$ collisions by the
number of detected wounded and evaporated neutrons.
Finally, to select $en + p_s$ events in $eD$ scattering a proton 
tagger with good $t$ resolution (to distinguish spectators from elastically
scattered protons) has to be built and installed at around $z=90$m.

\section{CONCLUSIONS}

HERA is so far a unique $lN$ collider and its potential must be 
fully exploited. It is clear that even with high precision HERA-2 data
a number of fundamental questions in QCD will remain open. HERA-3 
can provide new information for at least three items of
prime importance:
\begin{itemize}
 \item What is the partonic structure of the neutron at low $x$ and
       at large $Q^2, x$ ?
 \item What is the origin of confinement, and how can QCD describe 
       a regime of high parton density and weak coupling where
       cross sections ought to saturate?
 \item What is spin structure of the nucleon, especially at low $x$?  
\end{itemize} 
Since most of the infrastructure and the apparatus are in place
such a programme can be realized with relatively moderate investments
in three steps: $eD, ~\Dpol, ~eA (A=O^{16},Ca^{40})$ collisions.
The project is complementary to the planned future EIC facility
and may provide additional valuable information for the latter.


\begin{thebibliography}{99}
\bibitem{TESLA}  The TESLA project homepage, http://tesla.desy.de/
\bibitem{GLR}    L.V. Gribov, E.M. Levin and M.G. Ryskin, Phys.Rep. 100
                 (1981) 1.  
\bibitem{CGC}    E. Iancu, A. Leonidov and L. McLerran, SACLAY-T02/024, 
                 hep-ph/0202270 (2002)
\bibitem{THERA}  "The THERA Book", eds. U. Katz, M. Klein, A. Levy and 
                 S. Schlenstedt, DESY-01-123F, DESY-LC-REV-2001-062 (2001) 
\bibitem{EIC}    R. Milner, "A Future U.S. Electron-Ion Collider", these
                 proceedings; see also
                 http://www.phenix.bnl.gov/WWW/publish/abhay/Home\_of\_EIC/
\bibitem{HERMES} HERMES, A. Airapetian et al, Phys. Rev. Lett. 84 
                 (2000) 2584-2588.
\bibitem{GvS}    G. van der Steenhoven, "Spin Physics", the talk given in
                 \cite{WS3}.
\bibitem{WS1}    http://www.desy.de/heraea/ \\
                 "Physics with HERA as Electron-Nucleus Collider", eds.
                 G. Ingelman and M. Strikhman, May 1999; \\
                 "Future Physics at HERA", Proc. Workshop, DESY 1996/97,
                 eds. G. Ingelman, A. De Roeck and R. Klanner, pp.854-1092.
\bibitem{WS2}    "Physics with polarized protons at HERA", Proc. Workshop, eds.
                 A. De Roeck and T.Gehrmann, DESY-PROC-1998-01, Hamburg 1997;\\
                 "Polarized Protons at High Energies - Accelerator Challenges and
                 Physics Opportunities", Proc. Workshop, Hamburg 1999,
                 eds. A. De Roeck, D. Barber and G. R\"{a}del, DESY-PROC-1999-03,
                 Hamburg 1999.
\bibitem{WS3}    IPPP Workshop on Future Physics at HERA,
                 Durham, December 2001, hep-ex/0204032.
\bibitem{pol_D}  Ya.S. Derbenev and V.A. Anferov, 
                 Phys. Rev. ST - A\&B, 3, 094001 (2000)  
\bibitem{VFPS}   L. Favart, "Proposal for a Very Forward Proton
                 Spectrometer in H1 after 2000", DIS-2000,
                 hep-ph/0006167 (2000); 
                 P. Van Mechelen, "A Very Forward Proton Spectrometer for
                 H1", LISHEP-2002, hep-ex/0203029 (2002).
\bibitem{FNC}    V. Efremenko, "Forward Neutron Calorimeter for H1 
                 experiment at DESY", Proc. 9-th Int. Conf. CALOR 2000, 
                 Annecy, France (2000) 761.
\end{thebibliography}
\end{document}